\documentclass[letterpaper,twocolumn]{esapub} 
\usepackage{cranm_timesv}
\usepackage{epsfig}

\def\gtrsim{\mathrel{\hbox{\rlap{\hbox{\lower4pt\hbox{$\sim$}}}\hbox{$>$}}}}
\def\lesssim{\mathrel{\hbox{\rlap{\hbox{\lower4pt\hbox{$\sim$}}}\hbox{$<$}}}}

\title{\large
Coronal Heating versus Solar Wind Acceleration}

\author{Steven R. Cranmer}
\affil{Harvard-Smithsonian Center for Astrophysics,
Cambridge, MA 02138, USA; \,
{\em Email:} scranmer@cfa.harvard.edu \\
{\em To be published in the proceedings of}
SOHO--15: Coronal Heating, {\em 6--9 Sept.\  2004, St.\  Andrews,
Scotland, ESA SP--575}}

\begin{document}


\maketitle

\begin{abstract}
Parker's initial insights from 1958 provided a key causal
link between the heating of the solar corona and the
acceleration of the solar wind.
However, we still do not know what fraction of the solar
wind's mass, momentum, and energy flux is driven by
Parker-type gas pressure gradients, and what fraction is driven
by, e.g., wave-particle interactions or turbulence.
{\em{SOHO}} has been pivotal in bringing these ideas back to
the forefront of coronal and solar wind research.
This paper reviews our current understanding of coronal heating in
the context of the acceleration of the fast and slow solar wind.
For the fast solar wind, a recent model of Alfv\'{e}n wave
generation, propagation, and non-WKB reflection is presented
and compared with UVCS, SUMER, radio, and {\em in situ} observations
at the last solar minimum.
The derived fractions of energy and momentum addition from
thermal and nonthermal processes are found to be consistent
with various sets of observational data.
For the more chaotic slow solar wind, the relative roles of
steady streamer-edge flows (as emphasized by UVCS abundance
analysis) versus bright blob structures (seen by LASCO)
need to be understood before the relation between streamer
heating and and slow-wind acceleration can be known with
certainty.
Finally, this presentation summarizes the need for next-generation
remote-sensing observations that can supply the tight
constraints needed to unambiguously characterize the
dominant physics.
\end{abstract}

\section{Introduction}

The origin of coronal heating is intimately linked to the existence
and physical cause of the acceleration of the solar wind.
The early history of both ``unsolved problems'' reaches back
into the 19th century (e.g., Hufbauer 1991; Parker 1999, 2001;
Soon and Yaskell 2004).
Parker (1958, 1963) combined existing empirical clues concerning
an outflow of particles from the Sun with the earlier discovery
of a hot corona to postulate his transonic flow solution.
(An explicit closed-form solution to the isothermal Parker
solar wind equation was derived by Cranmer 2004.)
In Parker's original models, gravity was counteracted solely
by the large gas pressure gradient of the million-degree corona,
and wind speeds up to $\sim$1000 km/s were possible with
mean coronal temperatures of order 3--4 million K.

{\em Mariner 2} confirmed the existence of a continuous
supersonic solar wind just a few years after Parker's
initially controversial work, and also showed that the
wind exists in two relatively distinct states: slow
(300--500 km/s) and fast (600--800 km/s).
The succeeding decades saw a more comprehensive {\em in situ}
exploration of the solar wind.
Before the late 1970s, though, the slow-speed component of the
wind was believed to be the ``quiet'' background state
of the plasma; the high-speed streams were seen as occasional
disturbances (see Hundhausen 1972).
This view was bolstered by increasing evidence that average
coronal temperatures (in open magnetic regions feeding the solar
wind) probably did not exceed $\sim$2 million K, thus making
the slow wind easier to explain with Parker's basic theory.
However, we know now that this this idea came from the limited
perspective of spacecraft that remained in or near the ecliptic
plane; it gradually became apparent that the fast wind is indeed
the more ``ambient'' steady state
(e.g., Feldman et al.\  1976; Axford 1977).
The polar passes of {\em Ulysses} in the 1990s confirmed this
revised paradigm (Gosling 1996; Marsden 2001).

In the 1970s and 1980s, it became increasingly evident that
even the most sophisticated solar wind models could not
produce a {\em fast wind} without the deposition of
heat or momentum in some form into the corona
(e.g., Holzer and Leer 1980).
It is still unclear what fraction of the fast wind's
acceleration comes from the gas pressure gradient (i.e., from
coronal heating) and what fraction is directly added to the
plasma from some other source (usually believed to be waves).
This paper surveys our current understanding of
the fast wind with an eye on the relative impact of
coronal heating ({\S}~2) and external momentum deposition
({\S}~3).
A brief review of {\em SOHO} results concerning slow wind
acceleration---highlighting the similarities and differences
between the fast and slow wind---is given in {\S}~4.
Conclusions and a ``wish list'' of key measurements for future
missions are given in {\S}~5.

\section{Fast wind: Coronal heating}

Much of the SOHO--15 Workshop was devoted to studying the
so-called ``basal'' coronal heating problem; i.e., the
physical origin of the heat deposited below a heliocentric
distance of about 1.5 $R_{\odot}$.
At these heights, different combinations of mechanisms
(e.g., magnetic reconnection, turbulence, wave dissipation, and
plasma instabilities) are probably responsible for the varied
appearance of coronal holes, quiet regions, isolated loops, and
active regions (Priest et al.\  2000;
Aschwanden et al.\  2001; Cargill and Klimchuk 2004).
In the open magnetic flux tubes that feed the fast solar wind,
though, additional heating at distances greater than about
2 $R_{\odot}$ is believed to be needed (Leer et al.\  1982;
Parker 1991).
In coronal holes, the plasma at these larger heights is
almost completely collisionless.
Thus, the ultimate energy dissipation mechanisms at large
heights are probably {\em qualitatively different} from the
smallest-scale collision-dominated mechanisms (i.e., resistivity,
viscosity, ion-neutral friction) that act near the base.

The necessity for ``extended coronal heating'' in addition
to that at the base comes from three general sets of empirical
constraints (see also Cranmer 2002a).

\vspace*{-0.12in}
\begin{enumerate}
\item
As summarized above, pressure-driven models of the
high-speed wind cannot be
made consistent with the relatively low inferred temperatures in
coronal holes (especially electron temperatures $T_e$
less than about $1.5 \times 10^6$ K) without some kind of
additional energy deposition.
Because electron heat conduction is so much stronger than
proton heat conduction, it was realized rather early that
one cannot produce the observed {\em in situ} property of
$T_{p} > T_{e}$ at 1~AU without additional heating
(e.g., Hartle and Sturrock 1968).
\item
Spacecraft in the interplanetary medium have measured
radial gradients in proton and electron temperatures that
are substantially shallower than predicted from pure
adiabatic expansion, indicating gradual energy addition
(e.g., Phillips et al.\  1995; Richardson et al.\  1995).
{\em Helios} measurements of radial growth of the proton
magnetic moment between the orbits of Mercury and the
Earth (Schwartz and Marsch 1983; Marsch 1991) point to
specific collisionless processes.
\item
{\em SOHO} has provided more direct evidence for extended
heating.
UVCS (the Ultraviolet Coronagraph Spectrometer)
measured extremely high heavy ion temperatures, faster
bulk ion outflow compared to protons, and strong
anisotropies (with $T_{\perp} > T_{\parallel}$) of ion
velocity distributions in the extended corona
(Kohl et al.\  1997, 1998, 1999; Noci et al.\  1997;
Li et al.\  1998; Cranmer et al.\  1999b;
Giordano et al.\  2000).
SUMER (Solar Ultraviolet Measurements of Emitted Radiation)
has shown that preferential ion heating may begin very
near the limb, in regions previously thought to be in collisional
equilibrium and thus dominated by more traditional
heating mechanisms (e.g., Tu et al.\  1998;
Peter and Vocks 2003; Moran 2003; L.\  Dolla, these
proceedings).
\end{enumerate}

\vspace*{-0.05in}
The list of possible physical processes responsible for extended
coronal heating is limited both by the nearly collisionless
nature of the plasma and by the observed temperatures
($T_{\rm ion} \gg T_{p} > T_{e}$).
Most suggested mechanisms involve the transfer of
energy from {\em propagating fluctuations}---such as waves,
shocks, or turbulent eddies---to the particles.
This broad consensus has arisen because the ultimate
source of energy must be solar in origin, and thus it
must somehow be transmitted out to the distances where
the heating occurs (see, e.g., Hollweg 1978a;
Tu and Marsch 1995).
The {\em SOHO} observations discussed above have given rise
to a resurgence of interest in collisionless wave-particle
resonances (typically the ion cyclotron resonance)
as potentially important mechanisms for damping wave energy
and preferentially energizing positive ions
(e.g., McKenzie et al.\  1995;
Tu and Marsch 1997, 2001; Hollweg 1999a, 2000;
Axford et al.\  1999; Cranmer et al.\  1999a;
Li et al.\  1999; Cranmer 2000, 2001, 2002a,b;
Galinsky and Shevchenko 2000; Hollweg and Isenberg 2002;
Vocks and Marsch 2002; Gary et al.\  2003; Marsch et al.\  2003;
Voitenko and Goossens 2003, 2004;
Gary and Nishimura 2004; Gary and Borovsky 2004;
Markovskii and Hollweg 2004; see also E.\  Marsch, these
proceedings).

There remains some controversy over whether
ion cyclotron waves generated only at the coronal base
can heat the extended corona, or if a more gradual
generation of these waves is needed over a range of heights.
If the latter, then there is also uncertainty
concerning the origin of such extended wave generation.
MHD turbulence has long been proposed as a likely means of
transforming fluctuation energy from low frequencies (e.g.,
periods of a few minutes; believed to be emitted copiously
by the Sun) to the high frequencies required by cyclotron
resonance theories (e.g., 10$^2$ to 10$^4$ Hz).
However, both numerical simulations and analytic descriptions
of turbulence indicate that the cascade from large to
small length scales occurs most efficiently for modes that do
not increase in frequency (for a recent survey, see
Oughton et al.\  2004).
In the corona, the expected type of turbulent cascade would
tend to most rapidly increase electron $T_{\parallel}$, not
the ion $T_{\perp}$ as observed.
Cranmer and van Ballegooijen (2003) discussed this issue at
length and surveyed possible solutions.

Much of the work cited above can be broadly summarized as
``working backwards'' from the measured plasma parameters in
the extended corona to deduce the properties of the
kinetic-scale fluctuations that would provide the
required energy.
However, in many models (especially those involving turbulence)
the ultimate dissipation at small scales has its origin on
much larger scales.
It is therefore worthwhile to study the energy input at the
largest scales as a constraint on how much deposition
will eventually be channeled through the smaller scales.

The remainder of this section is thus devoted to presenting an
empirically constrained model of low-frequency (10$^{-5}$ to 1 Hz)
Alfv\'{e}n wave heating in a representative open coronal-hole
flux tube (Cranmer and van Ballegooijen 2004).
This model follows the radial evolution of the power spectrum of
non-WKB Alfv\'{e}n waves (i.e., waves propagating both outwards
and inwards along the flux tube) and allows the turbulent
energy injection rate (and thus the heating rate) to be derived
as a function of height.
The Alfv\'{e}n waves have their origin in the transverse shaking of
strong-field ($\sim$1500 G) thin flux tubes in the photosphere,
and in the supergranular network these flux tubes merge with one
another in the mid-chromosphere to form the bases of flux-tube
``funnels'' that expand outwards into the solar wind
(e.g., Hassler et al.\  1999; Peter 2001;
T.\  Aiouaz, these proceedings).

\begin{figure}
\centering
\epsfig{figure=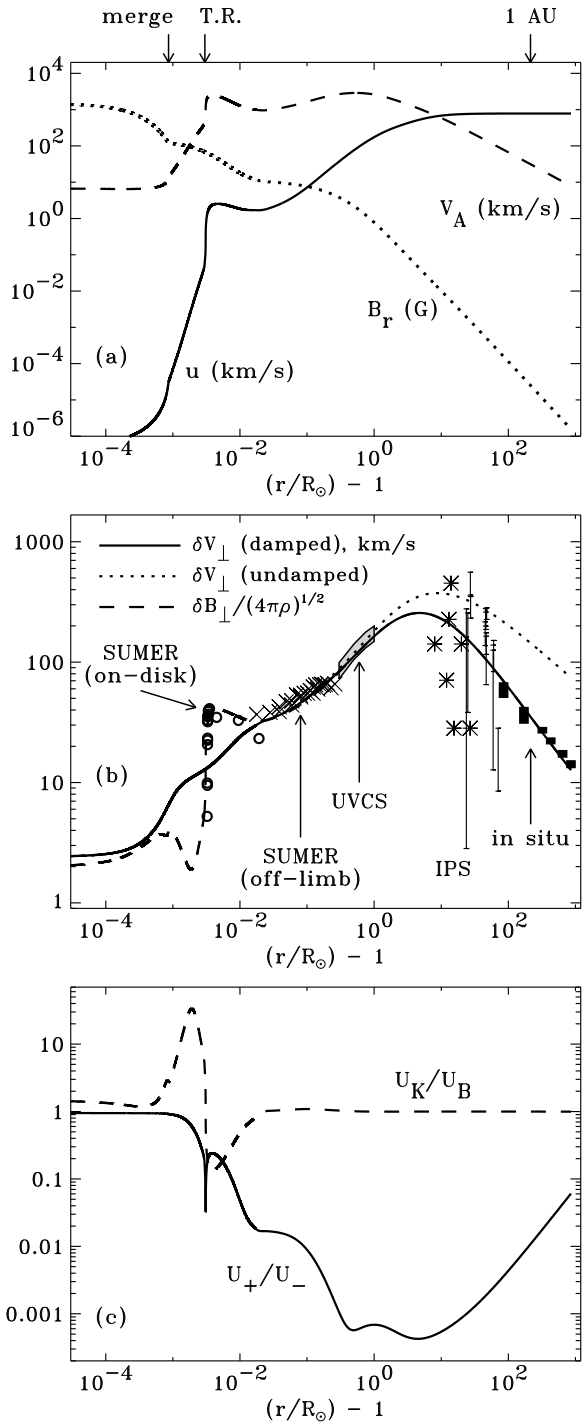,width=7.84cm}
\caption{{\bf (a)}
Steady-state plasma conditions along the modeled
flux tube: wind speed (solid line), Alfv\'{e}n speed
(dashed line), and magnetic field strength (dotted line).
Arrows at the top show the mid-chromospheric ``merging
height'' of thin flux tubes into network funnels,
the transition region, and the orbit of the Earth.
{\bf (b)}
Frequency-integrated wave amplitudes (see plot for line
styles).  Observational data points from left to right:
circles (Chae et al.\  1998),
X's (Banerjee et al.\  1998),
gray region (Esser et al.\  1999),
stars (Armstrong and Woo 1981),
struts (Canals et al.\  2002),
filled rectangles (Bavassano et al.\  2000).
{\bf (c)}
Energy density ratios defined in the plot.}
\end{figure}

Figure 1 shows a summary of various results from this model.
Figure 1a plots the adopted zero-order ``background'' plasma
state (magnetic field strength, wind speed, and Alfv\'{e}n speed)
on which the wave perturbations were placed.
The model extends from the photosphere into the outer
heliosphere (truncated at 4 AU for convenience).
The magnetic field $B_r$ was computed below 1.02 $R_{\odot}$ with a
2.5D magnetostatic model of expanding granular and supergranular
flux tubes (see, e.g., Hasan et al.\  2003).
Above 1.02 $R_{\odot}$, the magnetic field was adopted from
the solar-minimum model of Banaszkiewicz et al.\  (1998).
The density was specified empirically from VAL/FAL model C
(e.g., Fontenla et al.\  1993) at low heights, and white-light
polarization brightness measurements at large heights.
Mass flux conservation was used to compute the outflow speed,
normalized by the solar-minimum {\em Ulysses} polar mass flux
(for more details, see Cranmer and van Ballegooijen 2004).

The bottom boundary condition on the power spectrum of transverse
fluctuations came from measurements of G-band bright point motions
in the photosphere (e.g., Nisenson et al.\  2003).
The observationally inferred power spectrum was summed from two
phases of bright-point motion assumed to be statistically
independent: isolated random walks and occasional rapid jumps
due to flux-tube merging and fragmenting.
Below the mid-chromosphere, where the bright-point flux tubes
are isolated and thin, we solved a non-WKB form of the
kink-mode wave equations derived by Spruit (1981, 1984).
Above the mid-chromosphere, where the flux tubes have merged
into a more homogeneous network ``funnel,'' we solved the
wind-modified non-WKB wave transport equations of
Heinemann and Olbert (1980).
These wave equations were solved for each frequency in a grid
spanning periods from 3 seconds to 3 days, and the full
radially varying power spectrum was integrated to find the
kinetic and magnetic Alfv\'{e}n wave amplitudes $\delta V_{\perp}$
and $\delta B_{\perp}$.
Figure 1b shows these amplitudes for both the initial undamped
model and another model with turbulent damping (see below).
The various observational data points are described in the
caption.
We note here that the on-disk SUMER nonthermal line widths
of Chae et al.\  (1998) are most probably not transverse
Alfv\'{e}n waves, but their agreement with the {\em magnetic}
fluctuation amplitude in our model may imply some mode
coupling between transverse and longitudinal waves.
Figure 1c shows the departures from a simple WKB model of
purely outward-propagating Alfv\'{e}n waves.
Our model contains linear reflection that produces an
inward component of the wave energy density $U_{+}$ from
the predominantly outward component $U_{-}$ and does not
always exhibit the ideal WKB equipartition between
kinetic ($U_{K}$) and magnetic ($U_{B}$) fluctuations.
The total fluctuation energy density is given by
$U_{K} + U_{B} = U_{+} + U_{-}$.

Note that in Figure 1b the {\em in situ} measurements
fall well below the undamped wave amplitudes.
This heliospheric ``deficit'' of wave power, compared to
most prior assumptions about the wave power in the solar
atmosphere, is well known (Roberts 1989; Mancuso \& Spangler 1999).
It seems clear that damping is required in order to agree with
the totality of the measurements, and Cranmer
and van Ballegooijen (2004) showed that traditional collisional
(i.e., linear viscous) Alfv\'{e}n wave damping is probably
negligible in the fast solar wind.
However, if a turbulent cascade has time to develop, the
waves can be damped by small-scale kinetic/collisionless
processes at a rate governed by the large-scale energy
injection rate.
The most likely place for this damped wave energy to go is into
extended heating.

In a field-free hydrodynamic fluid, turbulent eddies are
isotropic and the energy injection rate
follows the Kolmogorov (1941) form.
This results in a volumetric heating rate (erg cm$^{-3}$ s$^{-1}$)
\begin{equation}
  Q_{\rm Kolm} \, \approx \,
  \frac{\rho \, \langle \delta V \rangle^3}{\ell}
\end{equation}
where $\rho$ is the mass density, $\langle \delta V \rangle$
is the r.m.s.\  fluctuation velocity at the largest scale
(called here, possibly imprecisely, the ``outer scale'')
and $\ell$ is a representative outer-scale length (i.e.,
the size of the largest turbulent eddies).
Heating rates of this general form were applied quite early
in studies of solar wind heating (Coleman 1968) and have
been used more-or-less continuously over the past few
decades (e.g., Hollweg 1986; Li et al.\  1999;
Chen and Li 2004).

In a magnetized low-beta plasma, the above Kolmogorov heating
rate does not apply because the turbulence is not isotropic.
In addition to the well-known MHD anisotropy that allows the
cascade to proceed much more efficiency in directions perpendicular
to the field than along the field, there is another (possibly
more important) departure from isotropy: the outward-propagating
Alfv\'{e}n waves (at outer-scale wavelengths) have a much
stronger amplitude than inward-propagating waves.
The outer-scale energy injection rate depends critically on
the disparity between the outward and inward wave energy
densities.
In terms of Elsasser's (1950) variables ($Z_{\pm} \equiv
\delta V \pm \delta B / \sqrt{4\pi\rho}$), where $Z_{-}$
represents outward waves and $Z_{+}$ represents inward waves,
the energy injection rate for anisotropic MHD turbulence can be
written as \begin{equation}
  Q \, = \, \alpha \, \rho \,
  \frac{\langle Z_{-}\rangle^{2} \langle Z_{+}\rangle +
        \langle Z_{+}\rangle^{2} \langle Z_{-}\rangle}
  {4 \ell_{\perp}}
\end{equation}
where $\alpha$ is an order-unity calibration factor and
$\ell_{\perp}$ is a purely transverse outer-scale
correlation length (see, e.g.,
Hossain et al.\  1995; Matthaeus et al.\  1999;
Dmitruk et al.\  2001, 2002).

In Figure 2 we plot the anisotropic and equivalent Kolmogorov
heating rates per unit mass ($Q/\rho$) for outer-scale
lengths that expand with the transverse width of the open flux
tube (i.e., $\ell_{\perp} \propto B_{r}^{-1/2}$).
The lengths are normalized so that the damping consistent with
the anisotropic heating rate matches the {\em in situ}
amplitudes in Figure 1b.
(The resulting normalization yields a value of $\ell_{\perp}$
at the chromospheric merging height of about 1100~km, which
seems appropriate for motions excited between granules of
the same spatial scale.)
We note that this model is completely consistent only above
$r = 1.1 \, R_{\odot}$, where both the damping and the heating
were computed together.  Below this height, Cranmer and
van Ballegooijen (2004) determined that the turbulence would
not have time to develop fully, and thus no damping was applied.
The heating rates below 1.1 $R_{\odot}$ should be considered
upper-limit estimates based on the undamped wave amplitudes.

\begin{figure}
\centering
\epsfig{figure=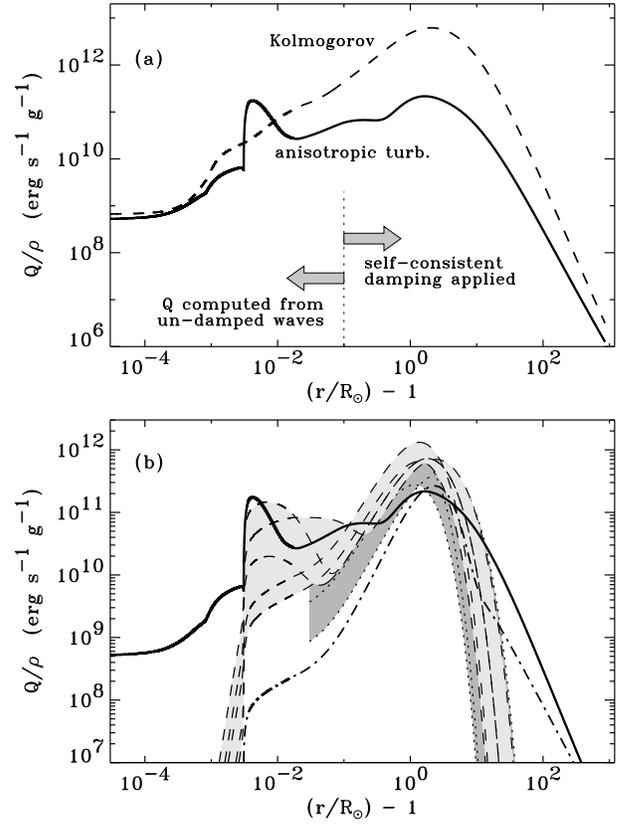,width=8.13cm}
\caption{{\bf (a)}
Heating rates per unit mass for the fully anisotropic MHD
turbulence model (solid line) and a model that assumes
isotropic Kolmogorov turbulence (dashed line).
{\bf (b)}
Comparing the same solid line from above with several sets
of empirically constrained heating rates:
dashed/light-gray (Wang 1994),
dotted/dark-gray (Hansteen and Leer 1995),
dash-dotted (Allen et al.\  1998).}
\end{figure}

Figure 2a shows the comparison between the anisotropic and
Kolmogorov heating rates.
The curves are substantially different from one another
nearly everywhere, which indicates that the inward/outward
imbalance generated by non-WKB reflection is probably a
very important ingredient in Alfv\'{e}n wave heating models
of the solar wind.
The differences are small in the photosphere and low
chromosphere, where strong reflection leads to nearly equal
inward and outward wave power.
In the extended corona, though, the Kolmogorov heating rate
begins to exceed the anisotropic turbulent heating rate by
{\em as much as a factor of 30.}
The isotropic Kolmogorov assumption assumes the maximal
amount of possible mixing between inward and outward modes,
which is inconsistent with the relatively weak reflection
computed for the corona in our models.

Figure 2b compares the derived anisotropic heating rate with
various empirically constrained heating rates---usually
specified via sums of exponential functions---from a selection
of 1D solar wind models.
In these models the parameters of these heating functions
were varied freely until sufficiently ``realistic'' solar wind
conditions were produced.
A selection of the models presented by Wang (1994) and
Hansteen and Leer (1995) are shown, and the SW2 model of
Allen et al.\  (1998) is plotted.
The order-of-magnitude agreement, especially in the extended
corona at $r \approx 1.5$--4 $R_{\odot}$, indicates that
MHD turbulence may be a dominant contributor to the extended
heating in the fast wind (see also Dmitruk et al.\  2002,
for similar comparisons).

\section{Fast wind: Direct Acceleration}

Just as electromagnetic waves carry momentum and exert
pressure on matter, acoustic and MHD waves that propagate through
an inhomogeneous medium also do work on the fluid via similar
radiation stresses.
This nondissipative net momentum deposition has been studied
for several decades in a solar wind context and is generally
called either ``wave pressure'' or a ponderomotive force
(e.g., Bretherton and Garrett 1968; Dewar 1970; Belcher 1971;
Alazraki and Couturier 1971; Jacques 1977).
Initial computations of the net work done on the bulk
fluid have been augmented by calculations of the acceleration
imparted to individual ion species
(Isenberg \& Hollweg 1982; McKenzie 1994; Li et al.\  1999;
Laming 2004),
estimates of the departures from Maxwellian velocity
distributions induced by the waves (Goodrich 1978;
Hollweg 1978b), and extensions to nonlinearly steepened wave
trains (e.g., Koninx 1992).

For non-WKB Alfv\'{e}n waves propagating along a radially
oriented (but potentially superradially expanding) flux tube,
Heinemann and Olbert (1980) gave the general expression for
the wave pressure acceleration $a_{\rm wp}$,
\begin{equation}
  \rho a_{\rm wp} \, = \,
  - \frac{\partial U_B}{\partial r} +
  \left( U_{B} - U_{K} \right)
  \frac{\partial}{\partial r} ( \ln B_{r} )
  \label{eq:awpHO80}
\end{equation}
where, as above, $U_B$ and $U_K$ are the magnetic and kinetic
energy densities of the waves,
\begin{equation}
  U_{B} = \frac{\langle \delta B_{\perp} \rangle^2}{8\pi}
  \,\,\, , \,\,\,\,\,\,\,\,\,
  U_{K} = \frac{\rho \langle \delta V_{\perp} \rangle^2}{2}
  \,\,\, .
\end{equation}
In the ideal WKB limit (i.e., for purely outward-propagating
Alfv\'{e}n waves), $U_{B} = U_{K}$ and only the
first term on the right-hand side is present.
The above expression also assumes an isotropic pressure
(i.e., $T_{\parallel} = T_{\perp}$ for the electrons and protons),
but for a low-beta plasma, modest departures from gas-pressure
isotropy do not substantially alter the wave pressure.
Cranmer and van Ballegooijen (2004) provide plots of $a_{\rm wp}$
versus height for the coronal hole flux tube model discussed
in {\S}~2.
We summarize those results briefly by mentioning that the weak
degree of reflection in the extended corona (leading to
$U_{B} \approx U_{K}$ above about 1.05 $R_{\odot}$) validates
the use of the simplified WKB form of the wave-pressure
acceleration in most solar wind models.

Rather than simply present plots of $a_{\rm wp} (r)$, here we
examine the impact of the ``known'' wave properties on the
acceleration region of the solar wind.
There are two semi-empirical ways of using the above-described
values for $\langle \delta V_{\perp} \rangle$ and $a_{\rm wp}$
to put constraints on the temperature of the extended corona.
Figure 3 shows coronal temperatures derived from the following
two methods:
\begin{enumerate}
\item
UVCS measurements of the widths of the H~I Ly$\alpha$
resonance line are useful for their sampling of the motions
of hydrogen atoms along the line of sight.
For the first few solar radii above the surface, efficient
charge exchange processes keep the proton and neutral hydrogen
temperatures coupled to one another.
For off-limb observations of coronal holes, the line of sight
samples mainly directions perpendicular to the nearly radial
field lines, and the $1/e$ line width $V_{1/e}$ arises from
two primary types of motion:
\begin{equation}
  V_{1/e}^{2} \, = \, \frac{2k_{\rm B} T_{p \perp}}{m_p} +
  \langle \delta V_{\perp} \rangle^{2}
\end{equation}
where $k_{\rm B}$ is Boltzmann's constant and $m_p$ is the
mass of a proton.
The two terms on the right-side represent random ``thermal''
motions and unresolved transverse wave motions.
Using observed values of $V_{1/e}$ and the modeled values of
$\delta V_{\perp}$, we can solve the above equation for
$T_{p \perp}$.
Note that the Cranmer et al.\  (1999b) data points in Figure 3a
were derived from $V_{1/e}$ values that have subtracted out
the projected component of the outflow speed along the line of
sight; the other values are straightforward line widths.
\item
If the steady-state density and outflow speed are known in
conjunction with the wave-pressure acceleration, the solar wind
momentum conservation equation can be solved empirically for
the gas pressure term:
\begin{equation}
  \frac{\nabla P}{\rho} \, = \, a_{\rm wp} -
  \frac{GM_{\odot}}{r^2} - u \frac{du}{dr}
\end{equation}
(see also Sittler and Guhathakurta 1999, 2002, for similar work).
To obtain the pressure $P$ as a function of radius, we integrated
$\nabla P$ inwards from 1 AU assuming a wide range of possible
temperatures at the outer boundary.
The resulting coronal $P(r)$ was quite insensitive to the boundary
conditions, however, because the gas pressure is so much larger
in the corona than at 1 AU.
An averaged proton-electron temperature $T_{\rm avg}$ was derived
assuming a fully ionized hydrogen-helium plasma:
\begin{equation}
  P \, = \, n_{p} k_{\rm B} T_{\rm avg} \left[
  2 + \frac{n_{\alpha}}{n_{p}} \left( 2 +
  \frac{T_{\alpha}}{T_{p}} \right) \right]
\end{equation}
where we assumed $n_{\alpha}/n_{p} = 0.05$ and we used two
extreme values for the alpha-to-proton temperature ratio:
1 and 4.
\end{enumerate}

\begin{figure}
\centering
\epsfig{figure=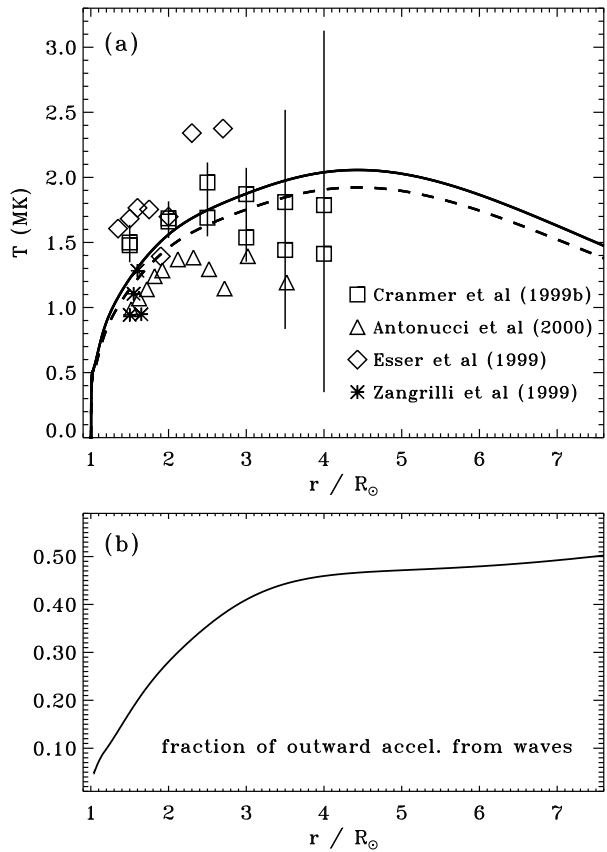,width=8.00cm}
\caption{{\bf (a)}
Derived coronal temperatures from UVCS H~I Ly$\alpha$
line widths (symbols), and from empirical momentum conservation.
For the latter, the two limiting values for the
alpha-to-proton temperature ratio $T_{\alpha}/T_{p}$ are
1 (solid line), and 4 (dashed line).
{\bf (b)}
Fraction of total fast-wind acceleration from wave pressure,
i.e., $\rho a_{\rm wp} / (\rho a_{\rm wp} + | \nabla P |)$.}
\end{figure}

Figure 3a displays the results of both kinds of semi-empirical
temperature determination discussed above.
The UVCS H~I Ly$\alpha$ line width observations (all in
solar-minimum polar coronal holes) exhibit a moderate spread
that probably can be attributed to different line-of-sight
contributions from polar plumes and low count-rate Poisson
statistics (as can be seen from the 1$\sigma$ error bars
plotted for the data points of Cranmer et al.\  1999b).
The overall agreement between both methods of determining the
temperature is an adequate consistency check, but note that
the UVCS-derived values are specifically {\em proton}
temperatures, while the momentum-conservation values are
essentially $(T_{p} + T_{e})/2$.
There is evidence from SUMER and CDS observations below
$r \approx 1.5 \, R_{\odot}$ that $T_e$ is substantially
less than 1 MK, and if this trend continues above 1.5
$R_{\odot}$, it would imply that the momentum-conservation
values of $T_p$ must be {\em larger} than plotted.
Thus, the rough agreement between the two methods in Figure 3a
may imply that $T_{p} \approx T_{e}$ remains the case several
solar radii out into the extended corona, in contrast with
earlier conclusions that $T_{p} > T_{e}$.

Figure 3b shows the fraction of the total outward acceleration
(gas pressure $+$ wave pressure) that comes from wave pressure.
This plot quantitatively answers the question that was implicitly
posed in the title of this paper; i.e., how do coronal heating
and direct acceleration ``compete'' in the fast solar wind?
The gas pressure term is decidedly stronger in the first
several solar radii (i.e., the primary fast-wind acceleration
region), but wave pressure soon reaches a point where it
provides roughly half of the acceleration.

All of the above discussion of wave-pressure acceleration was
focused solely on {\em Alfv\'{e}n waves,} but it is not yet
clear that these are the only MHD wave modes to exist in the
extended corona and solar wind.
There is some evidence for both fast-mode and slow-mode magnetosonic
waves in the corona, but they have been observed mainly in
relatively confined regions such as loops and plumes
(Ofman et al.\  1999; Nakariakov et al.\  2004).
Fast and slow modes are believed to be more strongly attenuated by
collisional damping processes than Alfv\'{e}n waves before they
reach the corona (e.g., Osterbrock 1961; Whang 1997).
However, fast-mode waves that propagate {\em parallel} to the magnetic
field behave essentially the same as Alfv\'{e}n waves (putting
aside their kinetic-scale polarization and their preferred
cascade directions in $k$-space), so they may exist at some low
level in the corona.

It is worthwhile, at least in a preliminary sense, to compare
the wave-pressure accelerations expected from Alfv\'{e}n waves
and from fast-mode waves.
For outward-propagating fast-mode waves with an isotropic distribution
of wavevectors, Jacques (1977) derived the especially simple expression
\begin{equation}
  \rho a_{\rm wp} \, = \,
  \frac{1}{3} \frac{\partial U_B}{\partial r} +
  \frac{4 U_B}{r}
\end{equation}
in the limit of zero plasma beta and $U_{K} = U_{B}$
(note also the opposite sign of the derivative term compared to
eq.~[\ref{eq:awpHO80}]).
The case of an isotropic distribution of wavevectors is likely
to be the case for fast-mode waves undergoing a turbulent cascade
(e.g., Cho and Lazarian 2003).
If we assume that Alfv\'{e}n and fast-mode waves have identical
amplitudes at $r = 2 \, R_{\odot}$, and that they both follow their
own linear wave-action conservation equations at heights
above and below 2 $R_{\odot}$, we can compare their respective
values of $a_{\rm wp}$ as a function of height.
Figure 4 shows this comparison, and above $r \approx 3 \, R_{\odot}$
the fast-mode acceleration is stronger than that of Alfv\'{e}n waves.
This is only an approximate and suggestive result, but it seems to
imply that a renewed study of fast-mode waves in the solar wind
is warranted (see also Habbal and Leer 1982;
Wentzel 1989; Kaghashvili and Esser 2000).

\begin{figure}
\centering
\epsfig{figure=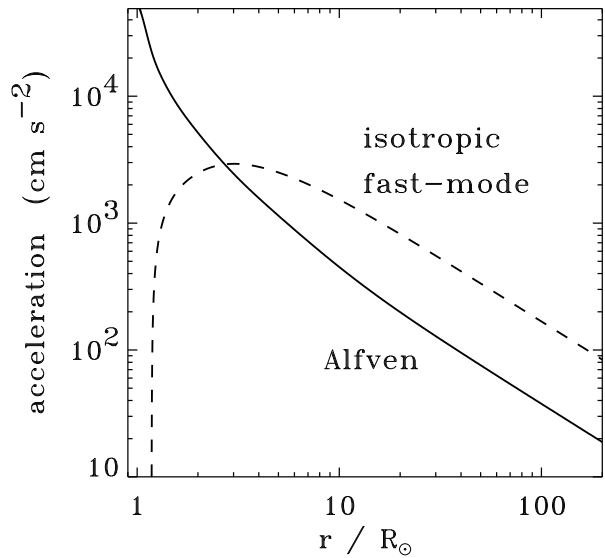,width=8.00cm}
\caption{Comparison of ideal WKB wave-pressure acceleration
for Alfv\'{e}n waves (solid line) and an isotropic
distribution of $\beta = 0$ fast-mode waves (dashed line).
Wave amplitudes were set equal to one another at $r=2 \, R_{\odot}$.}
\end{figure}

\section{Slow wind:$\,$ Similarities and \\ Differences}

The slow-speed component of the solar wind is believed to
originate mainly from the bright helmet streamers seen in
coronagraph images.
However, since these structures are thought to be mainly
closed magnetic loops or arcades, it is uncertain how the
plasma expands into a roughly time-steady flow.
Does the slow wind flow mainly along the open-field edges
of these closed regions, or do the closed fields occasionally
open up and release plasma into the heliosphere?
{\em SOHO} has provided evidence that both processes occur,
but an exact census or mass budget of slow-wind source
regions has not yet been constructed.
(This is a necessary prerequisite for studying slow-wind
``heating versus acceleration.'')

UVCS has shown, at least for the large quiescent equatorial band
at solar minimum, that streamers appear differently in the
emission of H~I Ly$\alpha$ and O~VI 1032 {\sc\aa}.
\linebreak[4]
The Ly$\alpha$ intensity pattern is similar to that seen in
LASCO visible-light images; i.e., the streamer is brightest
along its central axis.
In O~VI, though, there is a darkening in the core whose
only interpretation can be a substantial abundance depletion.
The solar-minimum equatorial streamers showed an oxygen
abundance of 0.3 the photospheric value along the streamer
edges, or ``legs,'' and between 0.01 and 0.1 times the
photospheric value in the core (Raymond et al.\  1997;
V\'{a}squez and Raymond 2004).
Low FIP (first ionization potential)
\linebreak[4]
elements such as Si and
Fe were enhanced by a relative factor of 3 in both cases
(Raymond 1999; see also Uzzo et al.\  2004).
Abundances observed in the legs are consistent
with abundances measured in the slow wind {\em in situ.}
This is a strong indication that the majority of the slow
wind originates along the open-field edges of streamers.
\linebreak[4]
The extremely low abundances in the streamer core, on the other
hand, are evidence for gravitational settling of the heavy
elements in long-lived closed regions, a result
that was confirmed by SUMER (Feldman et al.\  1998, 1999).

UVCS measurements have also been used to derive the wind
outflow speeds in streamers.
Strachan et al.\  (2002) found zero flow speed
at various locations inside in the closed-field core region
of an equatorial streamer.
Outflow speeds consistent with the slow solar wind were only
found along the higher-latitude edges and above the
probable location of the magnetic ``cusp'' between about
3.6 and 4.1 $R_{\odot}$.
Frazin et al.\  (2003) used UVCS to determine that O$^{5+}$
ions in the legs of a similar streamer have
significantly higher kinetic temperatures than hydrogen
and exhibit anisotropic velocity distributions with
$T_{\perp} > T_{\parallel}$, much like coronal holes
(see also Parenti et al.\  2000;
L.\  Strachan, these proceedings).
However, the oxygen ions in the closed-field core exhibit
neither this preferential heating nor the temperature anisotropy.
The analysis of UVCS data has thus led to evidence that
the fast and slow wind share at least some of the same
physical processes.

Evidence for another kind of slow wind in streamers came
from visible-light coronagraph movies.
The increased photon sensitivity of LASCO over earlier
instruments revealed an almost continual release of
low-contrast density inhomogeneities, or ``blobs,'' from
the cusps of streamers (Sheeley et al.\  1997;
see also Tappin et al.\  1999).
These features are seen to accelerate to speeds of order
300--400 km/s by the time they reach $\sim$30 $R_{\odot}$.
Wang et al.\  (2000) reviewed three proposed scenarios for
the production of these blobs:
(1) ``streamer evaporation'' as the loop-tops are
heated to the point where magnetic tension is overcome by
high gas pressure;
(2) plasmoid formation as the distended streamer cusp
pinches off the gas above an X-type neutral point; and
(3) reconnection between one leg of the streamer and an
adjacent open field line, transferring some of the trapped
plasma from the former to the latter and allowing it to
escape.
\linebreak[4]
Wang et al.\  (2000) concluded that all three mechanisms
might be acting simultaneously, but the third one seems to
be dominant.
Because of their low contrast, though
\linebreak[4]
(i.e., only about
10\% brighter than the rest of the streamer), the blobs
themselves cannot comprise a large fraction of the mass flux
of the slow solar wind.
This is in general agreement with the above abundance
results from UVCS.

Despite these new observational clues, the overall energy
budget in coronal streamers is still not well understood,
nor is their temporal MHD stability.
Recent models run the gamut from simple, but insightful,
analytic studies (Suess and Nerney 2002) to time-dependent
multidimensional simulations (e.g., Wiegelmann et al.\  2000;
Lionello et al.\  2001; Ofman 2004).
Notably, a two-fluid study by Endeve et al.\  (2004) showed
that the stability of streamers may be closely related to the
kinetic partitioning of heat to protons versus electrons.
When the bulk of the heating goes to the protons, the modeled
streamers become unstable to the ejection of massive plasmoids;
when the electrons are heated more strongly, the streamers are
stable.
It is possible that the observed (small) mass fraction of
LASCO blobs can give us an observational ``calibration'' of the
relative amounts of heat deposited in the proton and electron
populations.

\section{Conclusions and Future Missions}

Our understanding of the dominant physics of solar wind
acceleration has progressed rapidly in the {\em SOHO} era.
\linebreak[4]
Unfortunately, the multi-scale {\em complexity} of the plasma
in the extended corona has also been progressively revealed
during this same time period.
The solar physics community has benefited from increased
interaction with the space physics community, the latter having
decades more experience grappling with kinetic-scale plasma physics
and MHD turbulence.
It has been 5 years since Hollweg (1999b) asserted that the
``Holy Grail'' for
\linebreak[4]
theoreticians is the self-consistent modeling
of both the full wavenumber spectrum of MHD fluctuations and the
spatial dependence of proton, electron, and ion velocity distributions.
Much of the recent work cited in this paper, both observational
and theoretical, is helping the community get closer to this goal.


The remainder of this section highlights several areas where
future space missions (and future ground-based observatories
such as ATST) can provide key constraints that refine and test
theoretical explanations for solar wind acceleration.

The plasma parameters of both the major species (protons,
electrons, and He$^{2+}$) and minor ions are not yet known in
the wind's acceleration region with sufficient accuracy.
Figure 3a highlights the level of our uncertainty about $T_p$
and $T_e$ in coronal holes.
Progress in identifying some of the most basic aspects of
extended heating can be made only by constraining
these basic parameters more tightly.
In addition, only by better ``filling out'' our knowledge
of minor ion properties (as a function of ion charge and mass)
can we hope to uniquely identify the ultimate kinetic damping
mechanisms of waves and/or turbulence
(see Cranmer 2001, 2002b).
\linebreak[4]
{\em Spectroscopy is key}---especially in combination with
coronagraph occultation---in order to measure line profiles
out into the wind's acceleration region.

The full power spectrum of fluctuations (as a function of distance,
wavenumber $k_{\parallel}$ and $k_{\perp}$, and solar wind type)
is a strong driver of solar wind physics, but we still have
only indirect constraints on its properties in the corona.
The assimilation of multiple data sources, including radio sounding,
is crucial (e.g., Spangler 2002, 2003).
All previous {\em in situ} missions that measured wave power
spectra in the solar wind have been ``contaminated'' by the solar
rotation, which sweeps new, uncorrelated flux tubes past the
spacecraft on time scales of tens of minutes.
Cranmer and van Ballegooijen (2004) predicted that much of the
measured power with periods longer than about 30 minutes may be
due to this effect, and that a spacecraft that could sample the
fluctuations in a single flux tube would see intrinsically
higher-frequency ``fossil'' fluctuations from the Sun.
Solar {\em co-rotation} of
{\em in situ} missions (such as Solar
Orbiter) may be key, even if the co-rotation
is not exact or long-lived.

The origin of waves in jostled photospheric flux-tube
motions needs to be pinned down to a much better degree than at
present, in order to put firmer empirical constraints on the
``lower boundary condition'' of mechanical energy input into
the corona.
Synergy between 3D convection simulations and high-resolution
observations is becoming more common (e.g.,
S\'{a}nchez Almeida et al.\  2003).
Although space missions may one day boast collecting areas
rivaling those of ground-based telescopes, in the near future
it is the latter that will push the envelope to provide the
necessary constraints.  Existing sub-arcsecond spatial resolution
needs to be matched by sub-second time resolution, so that
the kinetic energy power spectra of small-scale flux tubes
(e.g., G-band bright point motions) can be measured more accurately.

\section*{Acknowledgments}

This work is supported by the National Aeronautics and Space
Administration (NASA) under grants NAG5-11913, NAG5-10996,
NNG\-04G\-E77G, and NNG\-04G\-E84G to the Smithsonian Astrophysical
Observatory, by Agenzia Spaz\-i\-ale Italiana, and by the
Swiss contribution to ESA's PRODEX program.

\end{document}